\documentclass[twocolumn,showpacs,preprintnumbers,amsmath,amssymb,prl]{revtex4}
\usepackage{amsmath}
\usepackage{dcolumn}
\usepackage{bm}
\usepackage{caption}
\usepackage{graphicx}

\newcommand*{\CHICAGO}{%
Enrico Fermi Institute, University of Chicago, Chicago, Illinois 60637, USA }

\newcommand{\kppnn}{K_{L}\rightarrow\pi^{0}\pi^{0}\nu\bar{\nu}}

\newcommand{\kppp}{K_{L}\rightarrow\pi^{0}\pi^{0}\pi^{0}}
\newcommand{\pion}{\pi^{0}}
\newcommand{\pt}{$P_{T}~ $}
\newcommand{\kpnn}{K_{L}\rightarrow\pi^{0}\nu\bar{\nu}}
\newcommand{\masspipi}{M_{\pion\text{-}\pion}}
\newcommand{\mevc}{\text{MeV}/\text{c}}
\newcommand{\mevcsq}{\text{MeV}/\text{c}^2}
\begin{document}

\title{Blind background prediction using a bifurcated analysis scheme}
\author{
J.~Nix, 
J. Ma,
G.N.~Perdue, 
Y.W.~Wah
\\}

\affiliation{
\CHICAGO
}
\date{\today}
\begin{abstract}
A technique for background prediction using data, but maintaining a closed signal box is described. The result is extended to two background sources. Conditions on the applicability under correlated cuts are described. This technique is applied to both a toy model and an analysis of data from a rare neutral kaon decay experiment.
\end{abstract}   
\pacs{06.20.Dk,14.40.Aq}
\maketitle

\newcommand\relphantom[1]{\mathrel{\phantom{#1}}}

In this paper we describe a bifurcation analysis procedure for data driven background prediction in a blind analysis of a closed signal region. In this type of analysis we define a signal region, in some set of variables, where we keep the events hidden during the analysis. We wish to predict the number of background events under the full set of cuts in this signal region. We use the term cut to refer to a specific selection criteria. An event that passes a cut is selected and an event which fails a cut is thrown out. The procedure uses the application of inverse cuts to properly measure the veto power of the different sets of cuts while keeping the signal region closed. This technique was first developed for a single background source in the stopped $K^+$ experiments E787 and E949 at Brookhaven \cite{e949}. The work in this paper was inspired by the use of the bifurcation technique in the E391a experiment \cite{e391}.

 We begin with a derivation of the bifurcation analysis in the case of one background source and uncorrelated cuts. We then extend this to two background sources and a simple model of correlation between cuts. Throughout this paper the method will be applied to a toy model to predict background and then applied to an example from the E391a experiment. We utilize the Mathematica software package to implement the toy model \cite{mathematica}.
 
 \section{One Background Case}\label{bkg1}
\begin{figure}[htbp]
  
   \includegraphics[width=.75\linewidth]{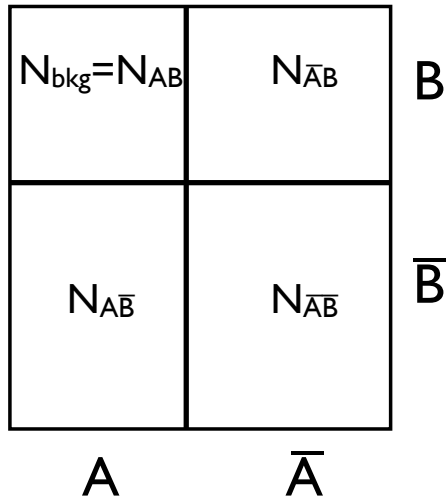} 
   \caption{Schematic of background distribution in the cut space.}
   \label{fig:geometricbifurcation}
\end{figure}

We begin discussing this method in the case of a single background source. Here a collection of setup cuts have been applied which eliminate all other sources of background. Our goal is to predict the amount of background in the signal region when we apply the cuts $A$ and $B$, which we refer to as the bifurcation cuts. The number of events we observe will be determined by the number of events before applying the cuts $A$ and $B$ (after applying the setup cuts) and the cut survival probability (CSP):
\begin{equation}
N_{\text{bkg}}=N_0   P(AB). 
\end{equation}
We consider events to lie in a multi-dimensional space with a dimension corresponding to every variable on which we can cut. Our set of cuts defines a multidimensional signal region which we wish to keep blind. If two cuts show no correlation in the events that they cut, this implies that these two cuts are orthogonal in this space. A diagram of this situation is shown in Fig. \ref{fig:geometricbifurcation}. The CSP can then be decomposed into $P(AB)=P(A)  P(B):$
\begin{equation}
N_{\text{bkg}}=N_0   P(A)  P(B). \label{eqn:nbkgdef}
\end{equation}
This can be expanded into
\begin{equation}
N_{\text{bkg}} = \frac{N_0^2   P(A)   P(B)   P(\bar{A})   P(\bar{B}) }{N_0   P(\bar{A})  P(\bar{B})}.
\end{equation}
Here $\bar{A}$ and $\bar{B}$ are the inverses of cuts $A$ and $B$, events which pass cut $A$ fail cut $\bar{A}$. Then we can calculate this from data based on the number of observed events in the signal region under the different cut conditions: 
\begin{align}
N_{A\bar{B}} = N_0   P(A)  P(\bar{B}), \label{eqn:nabbardef}\\ 
 N_{\bar{A}B}  = N_0   P(B)  P(\bar{A}), \\
N_{\bar{A}\bar{B}} = N_0   P(\bar{A})P(\bar{B}). \label{eqn:nabarbbardef}
\end{align}
Here $N_{A\bar{B}}$ is the number of background events observed with the application of cut $A$ and the inverse of cut $B$. $  N_{\bar{A}B}$ is the observed background events with the inverse of cut $A$ and cut $B$ applied. $N_{\bar{A}\bar{B}}$ is the count when the inverse of both $A$ and $B$ are applied. All of these values are outside the signal region defined in the multi-dimensional cut space allowing us to predict the background without opening the box:
\begin{equation}
N_{\text{bkg}} = \frac{ N_{A\bar{B}}    N_{\bar{A}B} }{N_{\bar{A}\bar{B}}}. \label{bkg1sol}
\end{equation}
The procedure for producing a background prediction is as follows. First, apply setup cuts to the data, the number of events in the signal box is $N_{0}$. The setup cuts are all the cuts other than $A$ or $B$, which are applied. They should remove all other background sources. There is more freedom in choosing the setup cuts than cuts $A$ and $B$, since they can be correlated with each other or $A$ and $B$. We then apply cut $A$ and $\bar{B}$. By applying $\bar{B}$, we are looking at events which are outside the signal box in the multidimensional space. We count the number of events which pass these sets of cuts, $N_{A\bar{B}}$. We then do the same procedure in reverse applying the set of cuts $B$ and the inverse of $A$ to find $N_{\bar{A}B}$.  Finally, we apply the inverse of both cuts $A$ and $B$ to find $N_{\bar{A}\bar{B}}.$ These values are combined to produce the background prediction.

Eqn \ref{bkg1sol} predicts no background events, if $N_{A\bar{B}}$ or $N_{\bar{A}B}$ are zero. This can be true for one of three reasons: $N_{0}=0$, $P(A)$ or $P(B)$=0, or $P(\bar{A})$ or $P(\bar{B})=0$. The first two possibilities are what we expect and reflect cases where there should be no background events. The third possibility is more problematic. When $P(\bar{A})$ or $P(\bar{B})=0$, $N_{\bar{A}\bar{B}}$ should also be zero, but statistical fluctuations may prevent that from being true. This condition results from a poor choice of cuts where one cut eliminates almost no background events. If possible a different choice of cuts for $A$ and $B$ should be made.


\section{The Toy Model}
Each event is described by four variables. Two kinematic variables, $p$ and $x$, which are used to describe the signal region and two cut variables, and $a$ and $b$, which will be used to define the cuts. The cut variables, $a$ and $b$, are independent of the kinematic variables, $p$ and $x$. All of these variables range from 0 to 1. 

\begin{figure}[htbp]
   \includegraphics[width=.9\linewidth]{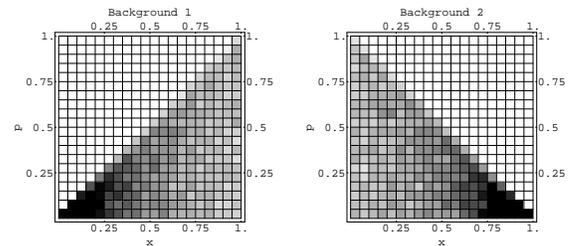} 
   \caption{Kinematic variable distributions for Backgrounds 1 and 2. }
   \label{fig:kinvar}
\end{figure}

We define 2 different types of events:  Background 1 and Background 2. They both have $x$ variables with uniform distributions between 0 and 1. Their $p$'s have uniform distributions between 0 and $x$ for Background 1 and between 0 and $1-x$ for Background 2. Background 1 has an uniform distribution of $a$ between 0 and 1 and variable $b$ has a linearly decreasing density with values between 0 and 1. Background 2's $a$ distribution is a linearly decreasing density with values between 0 and 1 and an uniform distribution between 0 and 1 for $b$. The distribution of each variable for the two background types is shown in Table \ref{variable}. The distributions of the kinematic variables $p$ and $x$ are shown in Fig. \ref{fig:kinvar}. We define a signal region by specifying the allowed kinematic variables: $0.25 < x < 0.75$ and $0.25 < p < 0.75$.
\begin{table}[htbp]
  \begin{tabular}{|l|c|c|}
\hline
Back- & $x$ & $p$  \\
ground & & \\
\hline
1 & $f(x)=1,~ x\in(0,1]$  & $f(p)=1/x,~ p\in[0,x]$  \\
& $f(x)=0,~ x\notin(0,1]$&$f(p)=0,~ p\notin[0,x]$ \\
\hline
 2 & $f(x)=1,~ x\in[0,1)$&  $f(p)=1/(1-x),~ p\in[0,1-x] $\\
& $f(x)=0,~ x\notin[0,1)$&$f(p)=0,~ p\notin[0,1-x]$ \\
\hline
\hline
Back- & $a$ & $b$ \\
ground & & \\
\hline
1 & $f(a)=1,~ a\in[0,1]$ & $f(b)=1-b,~ b\in[0,1]$ \\
  & $f(a)=0,~ a\notin[0,1]$& $f(b)=0,~ b\notin[0,1]$ \\
\hline
 2 &$f(a)=1-a,~ a\in[0,1]$ &  $f(b)=1,~ b\in[0,1]$ \\
& $f(a)=0,~ a\notin[0,1]$&   $f(b)=0,~ b\notin[0,1]$\\
\hline
\end{tabular}
 \caption{Probability distribution functions for the variables of each event type in the Toy Model.}
      \label{variable}
\end{table}

We define our cuts on variables $a$ and $b$ as:
\begin{align}
A = (a>0.5), \\
B = (b>0.5). 
\end{align}
$A$ and $B$ are true or false statements. If they are false the event is cut.  With the cut points defined, we can then calculate the CSP for each event type:  $P(A)$ or $P(B)$. In this toy model the CSPs can be calculated analytically because we know the underlying distributions. These values are shown in Table \ref{table:prob}.  
\begin{table}
\begin{tabular}{|l|c|c|}
\hline
Event Type & $P(A)$ & $P(B)$ \\
\hline
Background 1 & 0.5& 0.25\\
\hline
Background 2 &0.25 & 0.5\\
\hline
\end{tabular}
\caption{ Cut survival probabilities for each event type in the Toy Model.}
\label{table:prob}
\end{table}
\subsection{One Background in the Toy Model}
In this section we discuss the case of a single significant background. The background prediction is given by Eqn \ref{bkg1sol}. We generated $1\times10^{4}$ Background 1 events over the whole range of kinematic variables. This leaves $\approx 2200$ background events in the signal region before applying cuts $A$ and $B$. In Table \ref{singbkg}, we show the observed number of events for each combination of cuts, the predicted background, and the observed background after applying both cut $A$ and $B$. The predicted background of $267.8\pm20.6$ agrees well with the $256\pm16$ observed background events.
\begin{table}
\begin{tabular}{|l|c|}
\hline
$N_{0}$ & $2236\pm47$\\
\hline
$N_{A\bar{B}}$ &$831\pm29$ \\
\hline
$N_{\bar{A}B}$ & $280\pm17$\\
\hline
$N_{\bar{A}\bar{B}}$ &$869\pm29$ \\
\hline
Predicted Background &$267.8\pm20.6$ \\
\hline
Observed Background & $256\pm16$\\
\hline
\end{tabular}
\caption{Single background study for the Toy Model with only Background 1.}
\label{singbkg}
\end{table}

\section{Two Background Case}
\subsection{Derivation}
The previous derivation applied in the case of a single background source. If the background is made up of two different background sources, $N_0=N_1+N_2$, with different CSPs then there is a correlation introduced between cuts that must be accounted for. To do so, we begin by replacing Eqns \ref{eqn:nbkgdef} and \ref{eqn:nabbardef}-\ref{eqn:nabarbbardef} with: 
\begin{align}
N_{\text{bkg}}=N_1   P_{1}(A)  P_{1}(B) + N_2   P_{2}(A)  P_{2}(B), \label{eqn:nbkg2}\\ 
N_{A\bar{B}} = N_1   P_{1}(A)  P_{1}(\bar{B}) + N_2   P_{2}(A)  P_{2}(\bar{B}), \label{eqn:nabbar2}\\
 N_{\bar{A}B}  = N_1   P_{1}(\bar{A}) P_{1}(B) + N_2   P_{2}(\bar{A})  P_{2}(B), \label{eqn:nabarb2}\\
N_{\bar{A}\bar{B}} = N_1   P_{1}(\bar{A})P_{1}(\bar{B}) + N_2   P_{2}(\bar{A})  P_{2}(\bar{B}). \label{eqn:nabarbbar2}
\end{align}

Then our previous calculation of the background has a cross term introduced. We wish to find the correction to the one background solution. We begin by substituting the above definitions into the solution for the one background case, Eqn \ref{bkg1sol}:
\begin{equation}
\begin{split}
\frac{ N_{A\bar{B}}  N_{\bar{A}B} }{N_{\bar{A}\bar{B}}} 
&=\frac{1}{N_{\bar{A}\bar{B}}} \bigl[N_1   P_{1}(A) P_{1}(\bar{B}) + N_2  P_{2}(A) P_{2}(\bar{B})\bigr] \times \\
&\relphantom{=}{}\bigl[N_1  P_{1}(B) P_{1}(\bar{A}) + N_2  P_{2}(B) P_{2}(\bar{A}) \bigr].
\end{split}
\label{eqn:bkg2start}
\end{equation}
We expand the numerator:
\begin{equation}
\begin{split}
N_{A\bar{B}}  N_{\bar{A}B}
&= N_{1}^2 P_{1}(A)P_{1}(\bar{A})P_{1}(B)P_{1}(\bar{B}) \\
&\relphantom{=}{}+ N_{1}N_{2} \bigl[P_{1}(A)P_{2}(\bar{A})P_{2}(B)P_{1}(\bar{B}) \\
&\relphantom{=}{} +P_{2}(A)P_{1}(\bar{A})P_{1}(B)P_{2}(\bar{B}) \bigr] \\
&\relphantom{=}{} +N_{2}^2P_{2}(A)P_{2}(\bar{A})P_{2}(B)P_{2}(\bar{B}). \\ 
\end{split}
\end{equation}

We multiply $N_{\text{bkg}}$ by $N_{\bar{A}\bar{B}}$ to allow us to find the difference of Eqn \ref{eqn:bkg2start} and $N_{\text{bkg}}$:
\begin{equation}
\begin{split}
N_{\text{bkg}}N_{\bar{A}\bar{B}} 
&= \bigl[N_{1}P_{1}(A)P_{1}(B)+N_{2}P_{2}(A)P_{2}(B)\bigr] \times \\
&\relphantom{=}{}\bigl[N_{1}P_{1}(\bar{A})P_{1}(\bar{B})+N_{2}P_{2}(\bar{A})P_{2}(\bar{B}) \bigr] \\
&= N_{1}^2 P_{1}(A)P_{1}(\bar{A})P_{1}(B)P_{1}(\bar{B}) \\
&\relphantom{=}{} + N_{1}N_{2}\bigl[ P_{1}(A)P_{2}(\bar{A})P_{1}(B)P_{1}(\bar{B}) \\
&\relphantom{=}{}+ P_{2}(A)P_{1}(\bar{A})P_{2}(B)P_{1}(\bar{B}) \bigr] \\
&\relphantom{=}{} +N_{2}^2P_{2}(A)P_{2}(\bar{A})P_{2}(B)P_{2}(\bar{B}), \\
\end{split}
\end{equation}
\begin{equation}
\begin{split}
\frac{ N_{A\bar{B}}    N_{\bar{A}B} }{N_{\bar{A}\bar{B}}} 
&= N_{\text{bkg}}  \\
&\relphantom{=}{}+ \frac{N_{1}  N_{2}}{N_{\bar{A}\bar{B}}}   \biggl[  P_{1}(A)   P_{2}(\bar{A})   P_{2}(B)   P_{1}(\bar{B}) \\
&\relphantom{=}{}+ P_{2}(A)   P_{1}(\bar{A})   P_{1}(B)   P_{2}(\bar{B}) \\
&\relphantom{=}{}- P_{1}(A)   P_{2}(\bar{A})   P_{1}(B)   P_{2}(\bar{B}) \\
&\relphantom{=}{}- P_{2}(A)   P_{1}(\bar{A})   P_{2}(B)   P_{1}(\bar{B}) \biggr].
\end{split}
\end{equation}


The cross term vanishes if $P_{1}(A) =  P_{2}(A)$ and $P_{1}(B)=P_{2}(B)$, where for the purposes of the cuts the two backgrounds are the same.

We can simplify the cross term by rewriting the CSPs of the inverse cuts in terms of the CSPs of the cuts, $P_{i}(\bar{A}) = 1-P_{i}(A)$. Each element of the cross term has the same structure which can be expanded to: 
\begin{equation}
\begin{split}
P_{i}(A)P_{j}(\bar{A})P_{k}(B)P_{\ell}(\bar{B}) 
&= P_{i}(A)P_{k}(B)(1-P_{j}(A))\times\\
&\relphantom{=}{}(1-P_{\ell}(B))\\
&= P_{i}(A)P_{k}(B) \\
&\relphantom{=}{}- P_{i}(A)P_{j}(A)P_{k}(B) \\
&\relphantom{=}{}-P_{i}(A)P_{k}(B)P_{\ell}(B) \\
&\relphantom{=}{} +P_{i}(A)P_{j}(A)P_{k}(B)P_{\ell}(B) .\\
\end{split}
\end{equation}

Summing the elements of the cross term cancels everything except the terms with two CSPs:
\begin{equation}
\begin{split}
\frac{ N_{A\bar{B}}    N_{\bar{A}B} }{N_{\bar{A}\bar{B}}}
&= N_{\text{bkg}} \\
&\relphantom{=}{} + \frac{1}{N_{\bar{A}\bar{B}}} (N_{1}N_{2}(P_{1}(A)P_{2}(B)+P_{2}(A)P_{1}(B) \\
&\relphantom{=}{} -P_{1}(A)P_{1}(B)-P_{2}(A)P_{2}(B))\\
&=  N_{\text{bkg}} - \frac{N_{1}N_{2}(P_{2}(A)-P_{1}(A))(P_{2}(B)-P_{1}(B)}{N_{\bar{A}\bar{B}}}. \\
\end{split}
\end{equation}

We can further simplify the cross term by defining $\Delta_{A} = P_{2}(A)-P_{1}(A)$ and $\Delta_{B}=P_{2}(B)-P_{1}(B).$

\begin{equation}
N_{\text{bkg}} = \frac{ N_{A\bar{B}}    N_{\bar{A}B} }{N_{\bar{A}\bar{B}}} + \frac{N_{1}  N_{2}}{N_{\bar{A}\bar{B}}}  \Delta_{A}   \Delta_{B} \label{eqn:bkg2sol}
\end{equation}

The second term in Eqn \ref{eqn:bkg2sol} is not the contribution of a particular source to the background prediction. It is a correction to the prediction of the total number of background events from both sources.

\subsection{Properties of the Two Background Solution}
We now discuss the behavior of the Eqn \ref{eqn:bkg2sol}. This solution has the reasonable property that it is symmetric with respect to the definitions of the cuts $A$ and $B$ and the backgrounds 1 and 2. 

We consider the behavior of Eqn \ref{eqn:bkg2sol} under extreme conditions. First, we show that under no conditions can the total $N_{\text{bkg}}$ be negative.  The correction term will have it's maximum negative value when $\Delta_{A}=1$ and $\Delta_{B}=-1$ or $\Delta_{A}=-1$ and $\Delta_{B}=1$. Under these conditions, $N_{\bar{A}\bar{B}}=0$ and $N_{\text{bkg}}$ is undefined. We therefore want to study $N_{\text{bkg}}$'s behavior as we approach this limit. We begin by setting $\Delta_{B}=-1$ and studying the limit as $\Delta_{A}\rightarrow 1$. In this case cut $B$ removes Background 1 completely, but the Background 1 events which survive $\bar{A}B$ still contribute to the prediction produced by Eqn \ref{bkg1sol}.

The condition that $\Delta_{B}=-1$ sets what values the CSPs of the B can take:
\begin{align}
P_{1}(B) = P_{2}(\bar{B}) = 0, \\
P_{2}(B) = P_{1}(\bar{B}) = 1.
\end{align}
Substituting these values into Eqns \ref{eqn:nabbar2}-\ref{eqn:nabarbbar2}, we find: 
\begin{equation}
\begin{split}
N_{A\bar{B}}
&= N_{2} P_{2} (A),
\end{split}
\end{equation}
\begin{equation}
\begin{split}
N_{\bar{A}B}= N_{1} P_{1} (\bar{A}) = N_{1} ( 1 - P_{1}(A) ),
\end{split}
\end{equation}
\begin{equation}
N_{\bar{A}\bar{B}}= N_{2} P_{2} (\bar{A}) = N_{2} ( 1 - P_{2}(A) ).
\end{equation}

We then substitute these values into Eqn \ref{eqn:bkg2sol} and sum the two terms: 
\begin{equation}
\begin{split}
N_{\text{bkg}} 
&= \frac{N_{1} (P_{2}(A) - P_{2}(A)P_{1}(A) - P_{2}(A) + P_{1}(A))}{1-P_{2}(A)} \\
&= \frac{N_{1} P_{1}(A) (1-P_{2}(A))}{1-P_{2}(A)} = N_{1} P_{1}(A).
\end{split}
\end{equation}
As $\Delta_{A}\rightarrow 1$, $P_{1}(A) \rightarrow 0$. Therefore the $N_{\text{bkg}}$ goes to 0. This indicates that $N_{\text{bkg}}$ never has a non-physical negative value.

We now consider the case of $\Delta_{B}=0$. Here the  correction term is zero, but the contribution from the second background is not. We can see that the first term in Eqn \ref{eqn:bkg2sol} correctly predicts the background by simplifying Eqns \ref{eqn:nbkg2}-\ref{eqn:nabarbbar2}, substituting $P(B)$ for $P_{1}(B)$ and $P_{2}(B)$:
\begin{align}
N_{\text{bkg}}= (N_1P_{1}(A)+N_2 P_2(A))P(B),  \\
N_{A\bar{B}} = (N_1   P_{1}(A)+N_2   P_{2}(A) )P(\bar{B}), \\
N_{\bar{A}B}  = (N_1   P_{1}(\bar{A}) +N_2   P_{2}(\bar{A}) )P(B),  \\
N_{\bar{A}\bar{B}} = (N_1   P_{1}(\bar{A})+ N_2   P_{2}(\bar{A}) )P(\bar{B}).
\end{align}

Substituting these values into the first term of \ref{eqn:bkg2sol} and simplifying gives:
\begin{equation}
\frac{N_{A\bar{B}}N_{\bar{A}B}}{ N_{\bar{A}\bar{B}}}= (N_1P_{1}(A)+N_2 P_2(A))P(B).
\end{equation}
This is $N_{\text{bkg}}$, so the correction is not necessary to properly predict the background when one of the $\Delta$'s is zero.  

\subsection{Interpretation of the Two Background Solution} \label{sect:interpret2bkg}
It may seem counterintuitive that two backgrounds cannot be combined simply. This can best be understood as the second background introducing an implicit correlation. 

As an example, consider two backgrounds which individually have no correlation between cut $A$ and cut $B$, but do have different cut survival probabilities.  If $P_{1}(A)=0.75$ and $P_{1}(B)=0.75$, while $P_{2}(A)=0.25$ and $P_{2}(B)=0.25$ then the resulting combination of the two backgrounds will have a correlation. Events which survive cut $A$ are have a greater chance to survive cut $B$, because events which survive cut $A$ are more likely to be part of Background 1.  Events that don't survive cut $A$ are less likely to survive cut $B$, because they are more likely to be part of background 2. Therefore there is a correlation between cuts $A$ and $B$, even though for the individual backgrounds they are uncorrelated.

The values of $N_1$, $N_2$, $\Delta_A$, and $\Delta_B$ are not directly accessible in data without opening the signal box. There are two options: either derive these values from Monte Carlo or from other regions in signal space. Determining $N_{1}$ and $N_{2}$ generally will require both an alternative way of predicting one of the backgrounds and the value of $N_{0}$, the total number of background after setup cuts. This raises the question whether determining $N_{0}$ biases the analysis. From $N_{0}$ and the other observed background numbers, $N_{A\bar{B}}, N_{\bar{A}B},$ and $N_{\bar{A}\bar{B}},$ it is possible to effectively open the box and count $N_{\text{bkg}}$. Determining $\Delta_A$ and $\Delta_B$ also requires additional input. Their values can be derived from either Monte Carlo or data outside the signal region.

\subsection{Two Background Toy Model}
\begin{figure}[htbp]
   \includegraphics[width=.75\linewidth]{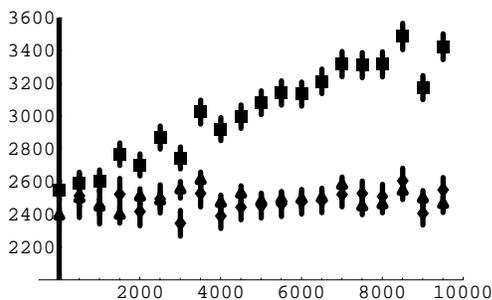} 
   \caption{Predicted and observed background for different admixtures of a Background 2. Squares are predicted background (without second background correction),  triangles are the observed background in data, diamonds are the corrected prediction. The x-axis is the number of generated Background 2 events, the total number of events, $N_{1}+N_{2},$ was held constant at $2\times10^{4}$.}
     \label{fig:normcor}
\end{figure}

In our toy model, we can study the effect of multiple background sources by varying the relative strength of a second background. We begin by calculating with the false assumption that there is a single background mode. We vary the relative admixture of Background 2. The total number of events, $N_{1}+N_{2},$ was held constant at $2\times10^{4}$. The discrepancy between the prediction and the observed background increases as the the number of background events from the second source increases, as shown in Fig. \ref{fig:normcor}. 

We now apply the correction term to the background prediction (Eqn \ref{eqn:bkg2sol}). In the case of this toy model, we know the values of $N_{1}$ and $N_{2}$ because we have set them. In a real analysis, it would be necessary to determine these values through either Monte Carlo studies or studies of different signal regions which are then extrapolated into the signal box. The differences in the cut probabilities, $\Delta_{A}$ and $\Delta_{B}$, also need to be determined from outside sources. In this model $\Delta_{A}=-25\%$ and $\Delta_{B}=25\%.$ Since the probability differences are of opposite signs the correction is negative and reduces the predicted background. 

In Fig. \ref{fig:normcor}, we show the results of keeping the total number of background events the same while increasing the fraction of Background 2 events. Here only the predicted background without correction increases, while the observed and corrected backgrounds remain relatively flat.

\section{Cut Correlation}
In the derivation of both the one and two background cases, we have assumed that the cuts $A$ and $B$ are uncorrelated. Of course, in real applications, it is unlikely to find two cuts which are perfectly uncorrelated. We would therefore like to find some general figure of merit to determine how correlation introduces errors into the background prediction. The following discussion will assume only one background source.

\subsection{Impact of Cut Correlation}
 To derive a correction to the background prediction, we use a simple model of cut correlation. We describe a case where the cuts have a weak linear correlation. In this model, the posterior probability for each cut is different than the prior probability. 
 

\begin{figure}[htbp]

    \includegraphics[width=.75\linewidth]{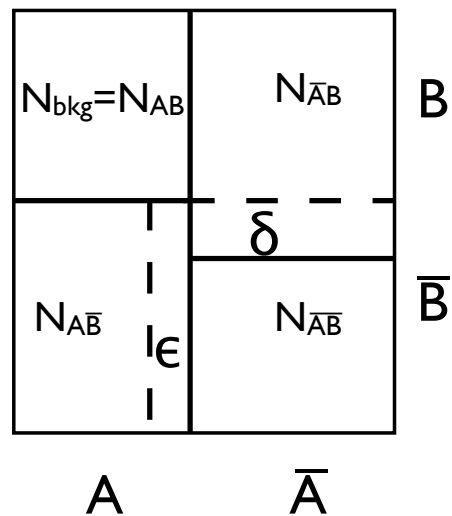} 

   \caption{Cut space with correlated cuts.}
   \label{fig:geometriccorrelation}
\end{figure}

We begin by specifying the background values in terms of the CSPs, which are now posterior probabilities, specifying the dependance on both cut conditions:
\begin{align}
N_{\text{bkg}} = N_0 P(A|B) P(B) = N_0 P(A) P(B|A), \\
N_{A\bar{B}} = N_0 P(A|\bar{B}) P(\bar{B}) = N_0 P(A) P(\bar{B}|A), \\
N_{\bar{A}B} = N_0 P(\bar{A}|B) P(B) = N_0 P(\bar{A}) P(B|\bar{A}), \\
N_{\bar{A}\bar{B}} = N_0 P(\bar{A}|\bar{B}) P(\bar{B}) = N_0 P(\bar{A}) P(\bar{B}|\bar{A}).
\end{align}

We proceed in the same fashion as for the two background case and substitute these definitions into the  solution (Eqn 7) for the single background, uncorrelated case: 
\begin{equation}
\frac{N_{A\bar{B}}N_{\bar{A}B}}{N_{\bar{A}\bar{B}}} = \frac{N_{0}^2 P(A|\bar{B})P(\bar{B})P(\bar{A}|B)P(B)}{N_0 P(\bar{A}|\bar{B}) P(\bar{B})}.
\end{equation}

We assume the correlations are small and relate the different posterior probabilities to each other: 
\begin{align}
P(A|\bar{B}) = P(A|B) - \epsilon, \\ \label{eqn:cordefinition1}
P(\bar{A}|\bar{B}) = P(\bar{A}|B) + \epsilon, \\ \label{eqn:cordefinition2}
P(B|\bar{A}) = P(B|A) - \delta, \\
P(\bar{B}|\bar{A}) = P(\bar{B}|A) + \delta.
\label{eqn:epsilondef}
\end{align}
The corrections $\epsilon$ and $\delta$ should be small. What we mean by small will be defined at the end of the derivation by what values are necessary for the corrections which are first order in $\epsilon$ and $\delta$ to be negligible. We have the freedom to choose to formulate the correction as a function of $\epsilon$ or $\delta$. We choose to use $\epsilon.$ We substitute the definitions into Eqn 31:

\begin{align}
\frac{N_{A\bar{B}}N_{\bar{A}B}}{N_{\bar{A}\bar{B}}} &= \frac{N_0 P(\bar{A}|B) (P(A|B)-\epsilon)P(B)}{P(\bar{A}|B)+\epsilon} \\
&= \frac{N_0 (P(A|B)P(B)-\epsilon P(B))}{1+\frac{\epsilon}{P(\bar{A}|B)}}.
\end{align}

Assuming the $\epsilon$ term in the denominator is small, we expand this result in $\epsilon$:
\begin{equation}
\begin{split}
\frac{N_{A\bar{B}}N_{\bar{A}B}}{N_{\bar{A}\bar{B}}} 
&\approx \bigl( N_0 (P(A|B)P(B)-\epsilon P(B))) \bigr) \\
&\relphantom{=}{} \times (1 - \frac{\epsilon}{P(\bar{A}|B)} + \frac{\epsilon^2}{P(\bar{A}|B)^2}+\mathcal{O}(\epsilon^3)).
\end{split}
\end{equation}
Multiplying this out and keeping the second order terms of $\epsilon$ gives:
\begin{equation}
\begin{split}
\frac{N_{A\bar{B}}N_{\bar{A}B}}{N_{\bar{A}\bar{B}}} 
&= N_{0}P(A|B)P(B) - \epsilon N_{0} \bigl(P(B)+ \frac{P(A|B) P(B)}{P(\bar{A}|B)}\bigr) \\
&\relphantom{=}{} + \epsilon^2 N_{0} \bigl(\frac{P(B)}{P(\bar{A}|B)} + \frac{P(A|B)P(B)}{P(\bar{A}|B)^2} \bigr). 
\end{split}
\end{equation}

The first term with no $\epsilon$ factors is $N_{bkg}$. The condition for the correlations to have a negligible impact on our background prediction is that the $\epsilon$ terms be much smaller than $\frac{N_{A\bar{B}}N_{\bar{A}B}}{N_{\bar{A}\bar{B}}}:$
\begin{equation}
\begin{split}
 N_{\text{bkg}}
&= \frac{N_{A\bar{B}}N_{\bar{A}B}}{N_{\bar{A}\bar{B}}} + \epsilon N_{0} P(B)\bigl(1+ \frac{P(A|B)}{P(\bar{A}|B)}\bigr) \\
&\relphantom{=}{} - \epsilon^2 N_{0} \frac{P(B)}{P(\bar{A}|B)} \bigl(1+ \frac{P(A|B)}{P(\bar{A}|B)} \bigr). 
\end{split}
\label{eqn:epscor}
\end{equation}

These terms require opening the signal box to know the correct values of the CSPs. We can however approximate these values with less knowledge, under the assumption that the number of events in the signal box is small: 
\begin{equation}
P(B)=\frac{N_{AB} + N_{\bar{A}B}}{N_{0}} \approx \frac{N_{\bar{A}B}}{N_{0}},
\end{equation}
\begin{equation}
\frac{P(A|B)}{P(\bar{A}|B)} = \frac{\frac{N_{AB}}{N_{B}}}{\frac{N_{\bar{A}B}}{N_{B}}} \approx \frac{N_{\text{pred}}}{N_{\bar{A}B}}.
\end{equation}

These approximations give us the first order correction:
\begin{align}
C_{\epsilon}= \epsilon  N_{\bar{A}B}(1+ \frac{N_{\text{pred}}}{N_{\bar{A}B}}). 
\label{eqn:practcor}
\end{align}

\begin{figure}[htbp]
   \includegraphics[width=.75\linewidth]{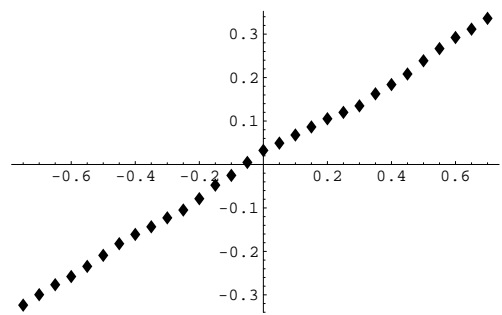} 
   \caption{The value of $\epsilon$ (Eqn \ref{eqn:cordefinition1}) as a function of $\epsilon'$ (Eqn \ref{adef}) in the Toy Model. }
   \label{fig:epsilon}
\end{figure}

Returning to our toy model, we introduce a correlation between the a and b variables in Background 1. We add a term linearly dependent on $b$ to $a$, and then scale $a$ to keep it between 0 and 1 and to reduce the change in background due to the change in the average value of $a:$ 
\begin{equation}
f(a)=(1+\epsilon'), a\in[\epsilon'b/(1+\epsilon'),(1+\epsilon'b)/(1+\epsilon)].
\label{adef}
\end{equation}
The variable $\epsilon'$ is the knob we use to tune the correlation. It is closely related to the variable $\epsilon$ that is defined in Eqns \ref{eqn:cordefinition1} and \ref{eqn:cordefinition2} as is shown in Fig. \ref{fig:epsilon}. 

\begin{figure}[htbp]
   \includegraphics[width=\linewidth]{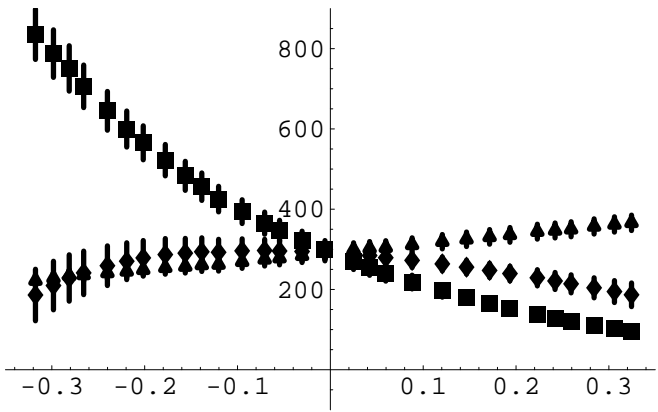} 
   \caption{Predicted and observed background for different levels of correlation between $a$ and $b$ in the Toy Model. Triangles are data background, squares are predicted, and diamonds are predicted  plus second order $\epsilon$ correction (Eqn \ref{eqn:epscor}). The x-axis is $\epsilon$.}
   \label{fig:corbkg}
\end{figure}

We show the predicted and observed background in Fig. \ref{fig:corbkg}.  As $\epsilon'$ increases the background in data increases, because the correlation increases the average value $a$, while the predicted background decreases. 


In Fig. \ref{fig:corbkg}, we show the prediction with $C_{\epsilon}$ added. It improves the agreement for a fairly wide range of $\epsilon$. 

\begin{figure}[htbp]
   \includegraphics[width=\linewidth]{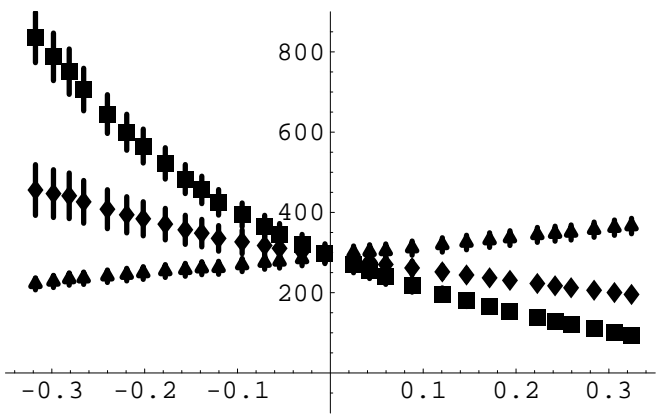} 
   \caption{Predicted and observed background for different levels of correlation between $a$ and $b$ in the Toy Model. Triangles are data background, squares are predicted, and diamonds are predicted plus practical $\epsilon$ correction (Eqn \ref{eqn:practcor}). The x-axis is $\epsilon$.}
   \label{fig:corbkgprac}
\end{figure}

\section{Example from E391a}
For a more realistic example, we discuss the use of this technique in the first search for the rare kaon decay $\kppnn$ at E391a. E391a is dedicated experiment for the search for the rare kaon decay $\kpnn$ located at KEK. The experimental apparatus consists of a $4\pi$ hermetic photon veto system and an array of CsI, an inorganic crystal scintillator, for signal detection. The $\kppnn$ decay has a Standard Model branching ratio of $(1.4\pm0.4)\times10^{-13}$ \cite{chiang}. The bifurcation method is to predict the background to this mode from data.

The final state of $\kppnn$ is four photons from the $\pion$'s and two unobservable $\nu$'s. Events are reconstructed by pairing the photons and calculating the $\pion$ decay vertices. The pairing with the smallest separation between $\pion$ vertices is selected. The signal region is defined in three kinematical variables: the transverse momentum of the $\pion$-$\pion$ system, the invariant mass of the the $\pion$-$\pion$ system, $\masspipi$, and reconstructed decay vertex, $Z$. 

\begin{figure}[htbp]
    \includegraphics[width=2.7 in,clip,angle=90]{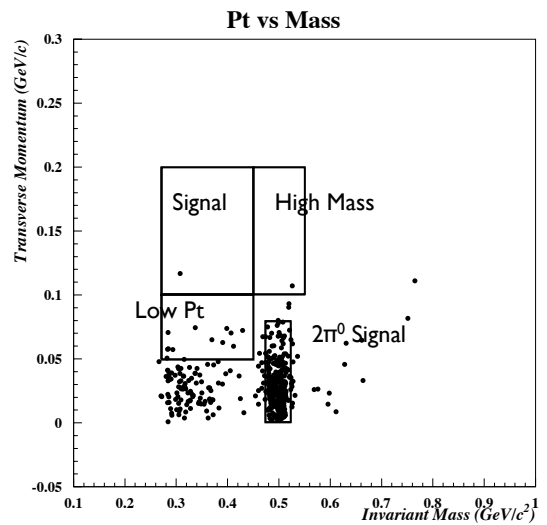}
   \caption{\pt plotted against mass with all cuts in the E391a experiment. $Z$ between 300 cm and 500 cm. The rectangular regions correspond to the regions in Table \ref{tbl:background} \cite{e391}. }
   \label{fig:ptmass}
 \end{figure}

The signal region is selected to avoid the two primary background sources: $\kppp$ decays and neutron interactions with a membrane located downstream of the fiducial region. The $\kppp$ events have a mass peak below the $K_{L}$ mass and with relatively low $P_T$'s of under $100 \mevc$. Their decay vertices cover the full range of the fiducial decay region. The neutron related events are well separated in the decay vertex having a peak at the membrane position, ~50 cm downstream of the fiducial region. The $\kppnn$ signal box is defined as $100 \mevc < P_{T} < 200 \mevc$, $268 \mevcsq< \masspipi < 400 \mevcsq$, $300 \text{cm} < Z < 500 \text{cm}$.  The distribution of events in the \pt-mass plane is shown in Fig. \ref{fig:ptmass}.

We divided our cuts into three groupings: the setup cuts, cut $A$ which contained most of the photon vetoes, and cut $B$ which contained primarily cuts on photon reconstruction quality in the CsI array. Using Eqn \ref{bkg1sol}, we first tested the method on regions surrounding the signal box. We define a Low \pt region with the same bounds on $\masspipi$ and $Z$ as the signal region and \pt $<100 \mevc$. The High Mass Region is defined with the same bounds on \pt and $Z$ as the signal region and $400 \mevcsq < \masspipi < 500\mevcsq$. The results of this are shown in Table \ref{tbl:background}. There is a significant discrepancy in the Low \pt region, this is due to neglecting the correlation.

\begin{table}[htb]
\caption{ Prediction of background events in different regions without correlation or secondary background corrections.}
\label{tbl:background}
\begin{tabular}{|c|c|c|c|c|c|}
\hline
 & & & & & \\
Region & $N_{\bar{A}\bar{B}}$ &$N_{A\bar{B}}$ & $N_{\bar{A} B} $ & Prediction & Data \\
\hline
Low \pt & $393$ & $72$ & $115$ & $21.1\pm3.3$ & 13  \\
\hline
High Mass & $46$ & $9$ & $4$ & $0.78\pm0.48$ & 1 \\
\hline
Signal & $84$ & $18$ & $2$ & $0.43\pm0.32$ & 1 \\
\hline
\end{tabular}
\end{table}

We estimated the cut correlation on the $\kppp$ background in its peak region using both data and Monte Carlo. The value of $\epsilon$ was found to be $(-.049\pm.035).$  We first apply Eqn \ref{eqn:practcor} on the Low \pt region and find a correction of $-6.67\pm4.81$. This correction brings the prediction in Table \ref{tbl:background} into much better agreement with data. We next apply  Eqn \ref{eqn:practcor} to the signal region and find a correction of $(-0.12\pm0.12)$ events.

Low energy neutron interactions are difficult to simulate in general and we had the additional difficulty of not knowing the precise shape of the membrane in the beamline. Therefore, we used data to estimate the level of neutron background. The value of $N_{2}$ was estimated by fitting the distribution events in the High \pt-High $Z$ region, $Z>500 \text{cm}$ and $P_{T}>100 \mevc$. Events in this region are predominately core neutron events. We applied a set of loose cuts to ensure high statistics. We subtracted off the contribution of the $\kppp$ background using a Monte Carlo prediction of their contribution under the loose set of cuts. The fitted core neutron distribution was extended into the signal region and the total number core neutron events in the signal region was predicted from the distribution. Then the predicted number of events in the signal region were scaled down using the factor by which the population of events in the High \pt-High $Z$ region were reduced. Using this procedure we estimate $N_{2}=2.16\pm0.03_{\text{stat.}}\pm1.05_{\text{syst.}}.$ The large systematic error comes from the subtraction of the $\kppp$ contribution in the High \pt-High $Z$ region. The differences between the cut survival probabilities for core neutron events and $\kppp$ events were found by comparing the difference in cut effectiveness between the $\kppp$ peak in the Low Mass-Low \pt region and the core neutron peak in the High Z region. The values which go into the correction and the computed correction are shown in Table \ref{tbl:coreneutron}.

\begin{table}[htb]
\begin{center}
\begin{tabular}{|c|c|c|c|c|}
\hline
$\Delta_{A}$ & $\Delta_{B}$ & $N_{1}$ & $N_{2}$ & Correction \\
\hline
26.4\% & 8.6\% & $(101.9\pm$&$(2.1\pm$ &$(0.06\pm0.05)$ \\
 & & $10.2\pm1.05)$ & $0.03\pm1.05)$ & \\
 \hline
 \end{tabular}
 \end{center}
 \caption{Values for the core neutron multiple background correction.}
 \label{tbl:coreneutron}
\end{table}

The $\kppp$ decay has three ways in which it can cause backgrounds depending on the number of photons lost through veto inefficiency or fusion of photon clusters in the CsI. These three channels are not separated in the signal space, so our estimate of cut correlation includes contribution from these multiple background types. 

The statistical uncertainties on both the correlation and the core neutron correction are large relative to their size. Therefore we decided to use these values as estimates of the systematic error made by neglecting the correlation and multiple background corrections. This results in a background prediction of $0.43\pm0.32_{\text{stat.}}\pm0.13_{\text{syst.}}$. Opening the signal box, we observed a single  event, consistent with our background prediction.

\section{Discussion}
The bifurcation analysis technique allows us to produce data driven background predictions while still maintaining a blind analysis. In this paper we have shown how to extend the bifurcation analysis to the case of two background sources and correlated cuts.

The correction for a second background source is effective for any level of secondary background. It does require information beyond which is available directly from a blind analysis. It requires a combination of Monte Carlo information about the relative strengths of the two backgrounds and how the cut survival probabilities vary between the two backgrounds. It also requires knowledge of $N_0$, the number of events in the signal box after the setup cuts. 

Correlations between the two cuts cannot be handled as easily. Even when the correlation between cuts is linear, the effects on the background prediction are non-linear. Therefore, special care must be taken when selecting the cuts for the bifurcation analysis to avoid correlation.

One particular aspect of the impact of the prediction on the cut correlation is it's dependance on the value of $N_{0}.$ This leads to two competing forces in optimizing the division of cuts into the setup cuts and the bifurcation cuts. From the perspective of minimizing statistical error in the bifurcation prediction, one would like a large value for $N_{0}$ with powerful cuts for cut $A$ and $B$, so that the statistical errors on the terms of Eqn \ref{bkg1sol} are small. On the other hand, a large $N_{0}$ means the prediction is sensitive to small correlations between the cuts. 

\bibliographystyle{plain}

\end{document}